\documentclass[preprint,12pt, 3p]{elsarticle}

\usepackage{amsmath}
\usepackage{amssymb}
\usepackage{mathrsfs}
\usepackage{hyperref}
\usepackage{bm}
\usepackage{accents}
\usepackage{xcolor}
\usepackage{graphicx}
\usepackage{wrapfig}

\newdefinition{rmk}{Remark}

\journal{arXiv}

\begin{document}

\begin{frontmatter}

\title{Evaluating fracture energy predictions using phase-field and gradient-enhanced damage models for elastomers}

\author[address1]{S. Mohammad Mousavi}
\author[address1,address2]{Ida Ang}
\author[address1,address3,address4]{Jason Mulderrig}
\author[address1,address5]{Nikolaos Bouklas\corref{corr}}
\ead{nb589@cornell.edu}
\address[address1]{Sibley School of Mechanical and Aerospace Engineering, \\ Cornell University, Ithaca, NY 14853, USA}
\address[address2]{Advanced Development, Mikel Inc. Middletown, RI 02842, USA}
\address[address3]{Materials and Manufacturing Directorate, Air Force Research Laboratory, Wright-Patterson Air Force Base, Dayton, OH 45433, USA}
\address[address4]{National Research Council (NRC) Research Associateship Programs, The National Academies of Sciences, Engineering, and Medicine, 500 Fifth St., N.W., Washington, DC, 20001}
\address[address5]{Center for Applied Mathematics, Cornell University, NY 14853, USA}

\cortext[corr]{Corresponding authors}

\begin{abstract}
Recently, the phase field method has been increasingly used for brittle fractures in soft materials like polymers, elastomers, and biological tissues. When considering finite deformations to account for the highly deformable nature of soft materials, the convergence of the phase-field method becomes challenging, especially in scenarios of unstable crack growth. To overcome these numerical difficulties, several approaches have been introduced, with artificial viscosity being among the most widely utilized.  This study investigates the energy release rate due to crack propagation in hyperelastic nearly-incompressible materials and compares the phase-field method and a novel gradient-enhanced damage (GED) approach. First, we simulate unstable loading scenarios using the phase-field method, which leads to convergence problems. To address these issues, we introduce artificial viscosity to stabilize the problem and analyze its impact on the energy release rate utilizing a domain J-integral approach giving quantitative measurements during crack propagation. It is observed that the measured energy released rate during crack propagation does not comply with the imposed critical energy release rate, and shows non-monotonic behavior. In the second part of the paper, we introduce a novel stretch-based GED model as an alternative to the phase-field method for modeling crack evolution in elastomers. It is demonstrated that in this method, the energy release rate can be obtained as an output of the simulation rather than as an input which could be useful in the exploration of rate-dependent responses, as one could directly impose chain-level criteria for damage initiation. We show that while this novel approach provides reasonable results for fracture simulations, it still suffers from some numerical issues that strain-based GED formulations are known to be susceptible to.
\end{abstract}


\end{frontmatter}

\section{Introduction}\label{Section:Intro}
Materials such as elastomers and hydrogels are frequently used in load-bearing applications due to their ability to sustain large deformations \cite{drury2003hydrogels, nonoyama2016double}. However, under high loading, these materials are prone to damage and fracture.  Damage and fracture can compromise the structural integrity of these materials, leading to ultimate failure and limiting their use in engineering and medical applications \citep{mark2003elastomers, sun2012highly, itskov2016rubber, tehrani2017effect, bai2017fatigue}. Consequently, a deeper understanding of how damage and fracture develop in soft materials is crucial for advancing industrial and novel engineering applications. Recent experiments have highlighted the complex cascade from damage to fracture in elastomers \cite{yang2019polyacrylamide, LIN2021101399, zhou2021flaw}, and several theoretical and computational works have aimed at addressing this phenomenon. It is also important to point out that the microstructural complexity\cite{lei2017influence, xiang2018general, chen2020mechanically} of these materials is often overlooked in macroscopic phenomenological models. This limits the predictive capabilities of these models, especially in the context of damage and fracture. Additionally, phenomenological constraints such as incompressibility or near-incompressibility increase the computational complexity of these problems.

In the computational engineering literature and practice, the phase-field method has been the \textit{modus operandi} for fracture simulations during the last decade. Considerable effort has been dedicated to the verification and validation of the phase-field method in the context of brittle fracture for linear elastic materials.  The variational formulation of brittle fracture, based on energy minimization, was first introduced by Francfort \textit{et al.} (1998) \cite{francfort1998revisiting} and later regularized by Bourdin \textit{et al.} (2000) \cite{bourdin2000numerical}. Unlike explicit methods like the Extended Finite Element Method (XFEM), which uses enriched basis functions to capture sharp discontinuities \cite{khoei2014extended, khoei2023modeling, khoei2023x}, the phase-field method represents cracks through a damage variable that transitions smoothly from zero in intact material to unity for fully damaged material. The phase-field method also introduces a length scale associated with the gradient of the damage variable for the regularization of the problem. This method avoids the explicit handling of discontinuities, representing interfaces as smooth transitions and eliminating the need for explicit tracking. 

Phase-field models for fracture mechanics, focusing on small strain and linear elasticity, have significantly advanced through various contributions to ensure proper algorithmic implementation \cite{miehe2010rate}, dynamic effects \cite{borden2012phase}, efficient hybrid formulations reducing computational costs \cite{ambati2015review}, and novel models integrating plastic strain-dependent fracture toughness \cite{yin2020ductile} (to name a few). These advancements collectively enhance the predictive accuracy and computational efficiency of phase-field simulations in fracture mechanics. Applying the phase-field method to soft materials including polymers and elastomers necessitates additional considerations, such as accommodating large deformations and addressing incompressibility. Significant efforts are also needed towards verification and validation of the phase-field theory predictions for elastomeric soft materials. This is a complicated task in itself due to the richness of the compositional landscape that needs to be explored. In a compressible setting where the Poisson ratio $\nu$ ranges from 0.3 to 0.45 \cite{miehe2014phase, raina2016phase, wu2016stochastic, tang2019phase, mandal2020length}, a pure displacement formulation is sufficient. However, for higher Poisson ratios, it is common to introduce a Lagrange multiplier (that we can identify as hydrostatic pressure field) to address the incompressibility constraint. This additional variable, combined with the displacement field, results in a mixed formulation of a saddle point problem. Solving this mixed formulation is prone to numerical issues due to difficulties in satisfying the inf-sup condition \cite{vassilevski1996preconditioning, benzi2005numerical, loghin2004analysis}. To prevent these issues, several numerical treatments have been proposed, including the use of Taylor-Hood spaces (above the lowest order)\cite{taylor2000mixed, onate2004finitecalculus, gavagnin2020stabilized, mang2020phase, alessi2020phase, suh2020phase, cajuhi2018phase, wriggers2021taylor} and various stabilization techniques \cite{klaas1999stabilized, maniatty2002higher, ang2022stabilized}. Specifically, Ang \textit{et al.} (2022) \cite{ang2022stabilized} introduced a mesh-dependent stabilization technique as an alternative to Taylor-Hood elements to address the inf-sup issue. Their study resulted in a model with lower computational costs and reduced residual pressures and stresses in the fully fractured state compared to previous studies \cite{tang2019phase, tian2020adaptive, swamynathan2021energetically}. In light of these computational and theoretical advancements, several recent works have successfully applied the phase field method to elastomer fracture situations \cite{arunachala2021energy, swamynathan2022phase, arunachala2023multiscale, feng2023phase, ye2023nonlinear, zhao2023phase, pranavi2024unifying}. In these studies, and more generally when utilizing the phase-field approach to brittle fracture, it is essential to provide the critical energy release rate as an input.  

An alternative approach to model fracture in elastomers and soft materials is the gradient-enhanced damage (GED) model paradigm \cite{de1996gradient, peerlings1996gradient, comi1999computational}. While this framework shares similarities with the phase-field model, such as employing a scalar damage variable, introducing a gradient term, and incorporating a length scale associated with the gradient, there is still a fundamental distinction between the two. Unlike the phase-field method, the gradient-enhanced damage model depicts material degradation via a damage evolution law reliant on the material's mechanical state, including state variables like the equivalent strain \cite{pham2010gradient, de2016gradient}. This is done so without considering the energy release rate as an input. Hence, the energy release rate can be derived as an outcome of this modeling process, which could prove beneficial compared to the phase-field method. Even so, gradient-enhanced damage models face a challenge with the broadening of the damage zone \cite{de2016gradient}, which is problematic especially when trying to emulate fracture. To address this, there has been a suggestion to introduce a functional dependence between the internal length scale parameter with the local strain or damage level \cite{geers1998strain}. 
Resolving the competition between localization from fracture and broadening of the damage zone is an open area for ongoing research, especially for strain-based GED formulations \cite{wosatko2021comparison, wosatko2022survey}. Specific advances towards accurate implementation of GED models for fracture, with damage as a primary unknown (and not equivalent strain), have been presented by \cite{lorentz2017nonlocal, talamini2018progressive}.
Gradient-enhanced damage models have primarily been used for modeling damage in materials like metals, concrete, and rocks \cite{pham2010gradient, kuhl2000anisotropic, marigo2016overview, seupel2019gradient, zhao2023modified}. Saji \textit{et al.} (2024) \cite{saji2024new} introduced a novel unified arc-length method that offers an alternative to the traditional Newton-Raphson approach for addressing convergence-related challenges during the material softening stage. Recently, the application of GED models for damage and failure in soft materials has become more appealing. Valverde-González et al. (2023) \cite{valverde2023locking} introduce two novel gradient-enhanced continuum damage formulations, Q1Q1E24 and Q1Q1P0, designed to tackle shear and volumetric locking issues in compressible and nearly incompressible hyperelastic materials.  Lamm \textit{et al.} (2024) \cite{lamm2024gradient1} proposed an extension for gradient-enhanced damage to model crack propagation in polymers undergoing large deformations. In their first paper, they presented a model that combines viscoelasticity with rate-dependent damage for finite deformations. To overcome the mesh dependencies of local damage, they introduced a global damage variable described by the micromorphic balance equation, as suggested by Forest (2009) \cite{forest2009micromorphic}, and solved this equation alongside the classical balance of linear momentum equation. In their subsequent work, Lamm \textit{et al.} (2024) \cite{lamm2024gradient2} extended this approach by incorporating a fully thermomechanically coupled material model, addressing the viscoelastic effects and damage within polymeric materials subject to finite strain.

This work focuses on modeling crack evolution in near-incompressible hyperelastic materials, focusing on verification of the results and consistency with the model parameters. The numerical stability of the phase-field method for unstable crack propagation, can lead to problems with convergence. To mitigate this issue, a common remedy is the incorporation of artificial viscosity penalizing the rapid evolution of the damage field. Here, we investigate the effect of artificial viscosity on the energy release rate \cite{miehe2010rate, miehe2014phase, sluys1992wave} as evaluated through a domain J-integral (and compare this result to the analogous model input). Furthermore, we introduce a new stretch-based GED model (that does not need the critical energy release rate as an input) as an alternative approach for modeling crack propagation in elastomers. We also examine how artificial viscosity influences this approach and how the results between the two methods compare. 

The outline of the paper is as follows: In Section \ref{Section:theory}, we outline the phase-field formulation for near-incompressible hyperelastic materials following \cite{ang2022stabilized}, summarize the strain-based GED, and introduce the new stretch-based GED along with the associated damage function. Section \ref{Section:NumericalImplementation} details the numerical implementation used in the open-source finite element platform \texttt{FEniCS} \cite{alnaes2015fenics}, and outlines the domain J-integral formulation for calculating the energy release rate. In Section \ref{Section:results}, we present five case studies that demonstrate the convergence characteristics of the solution, the mitigating effect of artificial viscosity, and the impact on the measured energy release rate. Finally, we furnish a comprehensive demonstration illustrating the results of the GED modeling, the effects of artificial viscosity on this method, and a comparison with the phase-field approach. 
\section{Formulation}\label{Section:theory}
\begin{figure}[h!]
    \centering
    \includegraphics[width=0.8\linewidth]{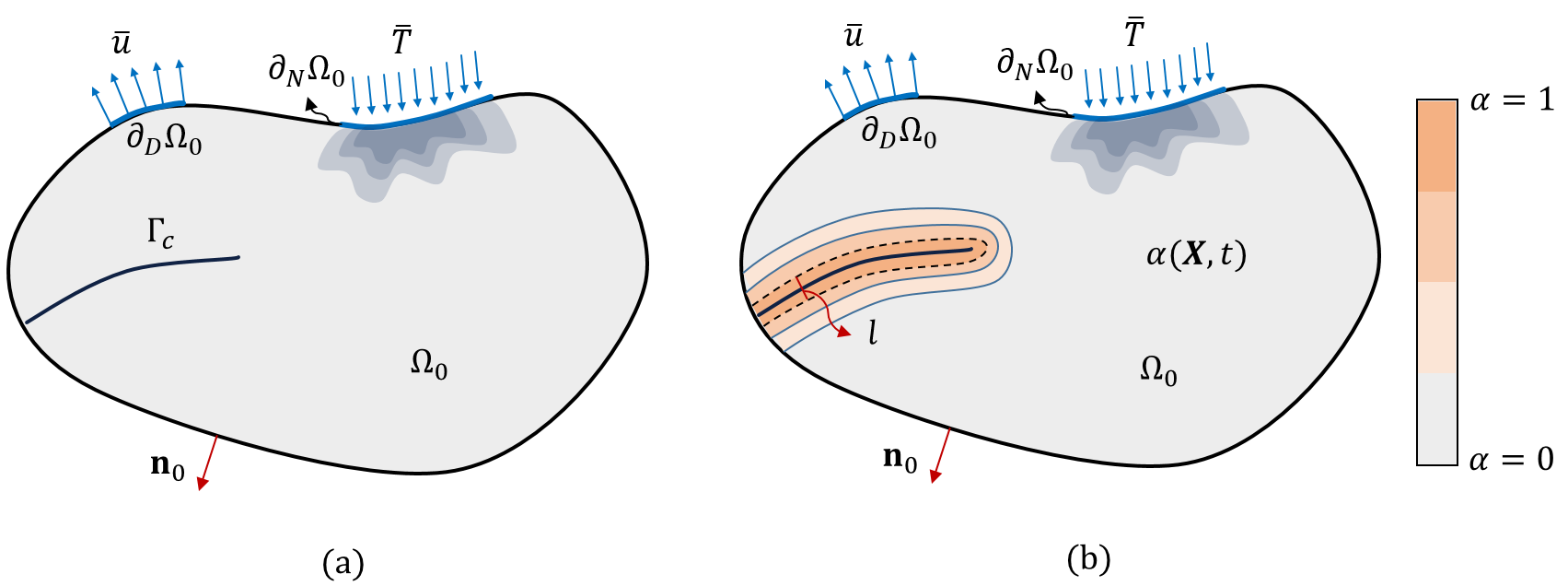}
    \caption{A continuum body in the reference configuration where a fracture is captured in a (a) discrete and (b) diffuse representation.}
    \label{fig:potato}
\end{figure}

Throughout all formulations and modeling procedures, the assumption is made that ${\Omega}_0 \subset \mathbb{R}^3$ represents an open, bounded, and connected subset, featuring a sufficiently smooth boundary denoted as $\partial \Omega_0$ (as depicted in Fig. \ref{fig:potato}).

\subsection{Kinematics}\label{SubSection:Kinematics}
As the body deforms, its constituent points follow a motion from their initial positions $\bm{X}$ in the reference configuration $\Omega_0$ to new positions $\bm{x}$ described in the current configuration $\Omega$. This transformation can be characterized by the displacement vector $\bm{u}$, given by $\bm{u} = \bm{x} - \bm{X}$, which indicates the motion of a specific particle from its original to its current position. However, to fully define the deformation, it is essential to delve into the kinematics of infinitesimal line elements, which can be elucidated using the following relationship:
\begin{equation}\label{bulk-element}
d\bm{x}=\textbf{F}(\bm{X},t)d\bm{X},
\end{equation}
where $\textbf{F}$ is the deformation gradient, defined as 
\begin{equation}\label{deformation-gradient}
\textbf{F}(\bm{X},t)=\frac{\partial\bm{x}}{\partial\bm{X}} \, .
\end{equation}
The Jacobian determinant of the deformation gradient, denoted as $J(\bm{X}, t)=dv/dV=det\textbf{F} (\bm{X},t)>0$, signifies the ratio between volume elements in the current and reference configurations. Additionally, we introduce the right Cauchy-Green deformation tensor $\textbf{C}$ as 
\begin{equation}\label{C}
\textbf{C}=\textbf{F}^T\textbf{F} \, ,
\end{equation}
along with its principal invariants \cite{holzapfel2002nonlinear}
\begin{equation}\label{invariants}
\begin{split}
I_1=\textrm{tr}(\textbf{C}), \\
I_2=\frac{1}{2}((\textrm{tr}(\textbf{C}))^2 - \textrm{tr}(\textbf{C}^2)), \\
I_3=\textrm{det}(\textbf{C}).
\end{split}
\end{equation}
\subsection{Hyperelasticity}\label{SubSection:Hyperelasticity}
We hereby consider a homogeneous, isotropic, and incompressible hyperelastic material, which experiences quasi-static loading. A Helmholtz free energy function $\mathcal{W}$ is introduced, defined per unit reference volume. 
The constitutive equation for hyperelastic materials thereby follows from the Coleman-Noll procedure as
\begin{equation}\label{constitutive}
\textbf{P}=\frac{\partial\mathcal{W}}{\partial\textbf{F}},
\end{equation}
where $\textbf{P}$ is the first Piola-Kirchhoff stress tensor. A variety of choices exist for the strain energy density function, from phenomenological and micromechanical models to data-driven models. Here, we will work with the compressible variant of the Neo-Hookean strain energy density function, commonly used for soft materials such as elastomers and tissues, and expressed as
\begin{equation}\label{neo-hookean}
{\mathcal{W}}(\mathbf{F}) =\frac{\mu}{2}(I_1(\mathbf{C})-3-2\ln J),
\end{equation}
where $\mu$ is the shear modulus. To enforce the near-incompressibility constraint, a perturbed Lagrangian formulation can be utilized \cite{ ang2022stabilized, brink1996some, li2020variational}. This involves introducing an additional variable, denoted as $p$, which acts as an indeterminate Lagrange multiplier. Through this approach, we can express the energy density function as follows

\begin{equation}\label{PF_strain_energy2}
\widetilde{\widetilde{\mathcal{W}}}(\mathbf{F},p) = \mathcal{W}(\mathbf{F}) -p\left(J -1\right) -\frac{p^2}{2\kappa},
\end{equation}
where $\kappa$ is a material parameter that controls the resistance to compressibility (similar to the bulk modulus) that should be sufficiently larger than the shear modulus to enforce near-incompressibility. Consequently, the constitutive equation \eqref{constitutive} can be specialized as

\begin{equation}\label{constitutive2}
\mathbf{P}= \frac{\partial\widetilde{\widetilde{\mathcal{W}}}(\mathbf{F},p)}{\partial \mathbf{F}} =\frac{\partial\mathcal{W}(\mathbf{F})}{\partial \mathbf{F}} - \, p \, J \, \mathbf{F}^{-T} \, .
\end{equation}

\subsection{Phase-field approximation of brittle fracture}\label{SubSection:PhaseField}

Within the phase-field method, a fracture is captured by smearing out a strong discontinuity in the displacement field which would otherwise represent a crack. This is accomplished through the consideration of a damage variable, denoted by a scalar field $\alpha: \Omega_0 \rightarrow [0, 1]$. As can be seen in Fig. \ref{fig:potato}(b), this field transitions from $0$, representing intact material to $1$, indicating fully damaged material \cite{bourdin2000numerical, pham2010gradient, marigo2016overview}. For cases at the near-incompressible limit, and following the recent implementations of Li \textit{et al.} (2020) \cite{li2020variational} and Ang \textit{et al.} (2022) \cite{ang2022stabilized}, we must adjust the strain energy density function to deteriorate as a function of the damage variable following

\begin{equation}\label{strain_energy2}
\widetilde{\mathcal{W}}(\mathbf{F},p,\alpha) = a(\alpha)\mathcal{W}(\mathbf{F}) -b(\alpha)p\left(J -1\right) -\frac{p^2}{2\kappa},
\end{equation}
where $a(\alpha)$ and $b(\alpha)$ are as follows

\begin{equation}\label{ab}
\begin{split}
a(\alpha) &= (1-k_{\ell})(1-\alpha)^2+k_{\ell}, \\
b(\alpha) &= (1-k_{\ell})(1-\alpha)^3+k_{\ell}.
\end{split}
\end{equation}
$k_{\ell}$ is a sufficiently small numerical conditioning parameter introduced for numerical stability. Following the aforementioned works, it is noteworthy that $b$ exhibits a higher polynomial order than $a$, which ensures that the effective resistance to bulk deformations deteriorates faster than the resistance to shear. This design ensures that the incompressibility constraint does not impede the physical opening of cracks. 

Furthermore, we can reformulate the constitutive equation using the adjusted energy density function \eqref{strain_energy2}, which has been modified to consider the impact of damage, in the following manner

\begin{equation}\label{constitutive3}
\mathbf{P}= \frac{\partial \widetilde{\mathcal{W}}(\mathbf{F},p,\alpha)}{\partial \mathbf{F}} =a(\alpha)\frac{\partial\mathcal{W}(\mathbf{F})}{\partial \mathbf{F}} - b(\alpha) \, p \, J \, \mathbf{F}^{-T}.
\end{equation}

The total energy of the system can be expressed as a functional 
\begin{equation}\label{energy_functional0}
\mathcal{\psi} \left(\bm{u}, p\right)=  
\int_{\Omega_0} \widetilde{\widetilde{\mathcal{W}}}(\mathbf{F},p) \, dV +\mathcal{G}_c\int_{\Gamma_c}d\Gamma -\int_{\partial_N\Omega_0}{\bm{T}}_0\cdot\bm{u}\, d A,
\end{equation}
where $\mathcal{G}_c$ represents the critical energy release rate and $\Gamma_c$ the crack area, as shown in Fig. \ref{fig:potato} (a).
Finally, by using this degraded strain energy density function, the total energy of the system can be approximated as \cite{miehe2010thermodynamicPF} 
\begin{equation}\label{energy_functional1}
\mathcal{\psi} \left(\bm{u}, p, \alpha\right) \approx 
\int_{\Omega_0} \widetilde{\mathcal{W}}(\mathbf{F},p,\alpha) \, dV +\frac{\mathcal{G}_c}{c_w}\int_{\Omega_0} \left(\frac{w(\alpha)}{\ell} + \ell \Vert \nabla \alpha \Vert^2 \right) dV -\int_{\partial_N\Omega_0}{\bm{T}}_0\cdot\bm{u}\, d A,
\end{equation}
where $\mathcal{G}_c$ represents the critical energy release rate. Meanwhile, $\ell>0$ is the length parameter and serves as a numerical regularization. Its value defines the width of the diffuse crack and acts as an internal length within the model \cite{pham2010gradient, marigo2016overview}. The term $c_w = \int_0^1\sqrt{w(\alpha)}d\alpha$ functions as the normalization constant, where $w(\alpha)=\alpha$ denotes an increasing function signifying specific energy dissipation per unit volume. The traction force, denoted by $\bm{T}_0$, operates on the Neumann boundary $\partial_N\Omega_0$ of the domain $\Omega_0$.

\subsubsection{Weak and strong forms}
In the context of variational fracture mechanics relating to the phase-field model, the damage progression in a continuum body is regulated by three main principles: irreversibility, stability, and energy balance \citep{pham2010gradient,marigo2016overview,baldelli2021numerical}. According to the first-order stability condition,
\begin{equation}
d\mathcal{\psi} \left(\bm{u}, p, \alpha; \, \bm{v}, q, \beta\right) \ge 0, \quad \forall (\bm{u}-\Bar{\bm{u}}, p, \alpha) \in (\mathbb{U}, \mathbb{P}, \mathbb{V}), \; \forall (\bm{v}, q, \beta) \in (\mathbb{U}, \mathbb{P}, \mathbb{V}),
\end{equation}
where $d\mathcal{\psi} \left(\bm{u}, p, \alpha; \bm{v}, q, \beta\right)$ represents the Gâteaux derivative of the functional \eqref{energy_functional1} evaluated at $(\bm{u}, p, \alpha)$ in the direction of the variations $(\bm{v}, q, \beta)$. The function spaces $\mathbb{U}$, $\mathbb{P}$, and $\mathbb{V}$, are defined in references \citep{brink1996some,le1994numerical} and \citep{valverde2023locking}, respectively. Hence, we obtain the following weak form:
\begin{equation}\label{weak_form_phase_field}
\begin{split}
\int_{\Omega_0} \mathbf{P}  : \nabla\bm{v} \,dV -\int_{\partial_N\Omega_0}{\bm{T}}\cdot\bm{v} \,dA &=0,\\
\int_{\Omega_0}  \left(b(\alpha)\left(J-1\right) -\frac{p}{\kappa}\right)q \,dV &=0,\\ 
\int_{\Omega_0}  \frac{\partial \widetilde{\mathcal{W}}(\mathbf{F},p,\alpha)}{\partial \alpha}\beta 
\,d V +\frac{\mathcal{G}_c}{c_w\ell}\int_{\Omega_0} \left(\frac{\partial w(\alpha)}{\partial \alpha} \beta  + 2\ell^2 \nabla \alpha\cdot \nabla \beta  \right) dV &=0.
\end{split}
\end{equation}
Thus, by Eq. \eqref{PF_strain_energy2}, the strong form is
\begin{equation}\label{strong_forms_phase_field}
\begin{split}
\nabla\cdot \mathbf{P} - \bm{T} &=0 \quad \;\;\text{in} \quad {\Omega}_0,\\
- b(\alpha)(J-1) -\frac{p}{\kappa}&=0 \quad \;\;\text{in} \quad {\Omega}_0,\\
\frac{\partial\widetilde{\mathcal{W}}\left(\mathbf{F},p, \alpha\right)}{\partial\alpha} +\frac{\mathcal{G}_c}{c_w\ell}\left(\frac{\partial w(\alpha)}{\partial\alpha} - 2\ell^2\nabla^2\alpha\right) &=0 \quad\;\; \text{in} \quad {\Omega}_0,\\
\end{split}
\end{equation}
along with the following boundary conditions
\begin{equation}\label{BC_phase_field}
\begin{split}
\bm{u} &= \Bar{\bm{u}} \quad \text{on} \quad \partial_{D}{\Omega}_0,\\
{\bm{T}} &= \Bar{\bm{T}} \,\quad \text{on} \quad \partial_{N}{\Omega}_0,\\
\alpha &= \Bar{\alpha} \quad \text{on} \quad \partial_{D}{\Omega}_0,\\
\bm{n}_0.\nabla\alpha &= 0 \,\quad \text{on} \quad \partial_{N}{\Omega}_0,\\
\end{split}
\end{equation}
where $\bm{n}_0$ is outward normal to the boundary, and $\Bar{\bm{u}}$ indicates a prescribed Dirichlet boundary condition on the complementary boundary part $\partial_{D}{\Omega}_0$ (as shown in Fig. \ref{fig:potato}).

As mentioned earlier, to overcome numerical stability issues due to unstable crack growth, the use of artificial viscosity is commonly used in phase-field formulations in order to penalize the rapid growth of the damage field \cite{ulloa2022variational, aldakheel2021multilevel}. Proposed by Miehe \textit{et al.} (2010) \cite{miehe2010rate, miehe2010thermodynamicPF}, we can adapt our current formulation through the strong form $\eqref{strong_forms_phase_field}_3$ as follows
\begin{equation}\label{eq_viscosity}
\frac{\partial\widetilde{\mathcal{W}}\left(\mathbf{F},p, \alpha\right)}{\partial\alpha} +\frac{\mathcal{G}_c}{c_w\ell}\left(\frac{\partial w(\alpha)}{\partial\alpha} - 2\ell^2\nabla^2\alpha\right) + \eta\Dot{\alpha}=0 \quad\;\; \text{in} \quad {\Omega}_0,\\
\end{equation}
where $\eta$ is the artificial viscocity coefficient and $\dot{(\,\,)}$ indicates the material time derivative (following the nomenclature in \cite{miehe2010rate}). In accordance with this change, $\eqref{weak_form_phase_field}_3$ from the weak form also has to be modified as
\begin{equation}\label{weak_form1}
\int_{\Omega_0}  \frac{\partial \widetilde{\mathcal{W}}(\mathbf{F},p,\alpha)}{\partial \alpha}\beta 
\,d V +\frac{\mathcal{G}_c}{c_w\ell}\int_{\Omega_0} \left(\frac{\partial w(\alpha)}{\partial \alpha} \beta  + 2\ell^2 \nabla \alpha\cdot \nabla \beta  \right) dV + \int_{\Omega_0} \eta\Dot{\alpha}\beta dV=0.
\end{equation}

\subsection{Gradient-enhanced damage model}\label{SubSection:GED}
Even though the phase-field method provides a robust framework for predictive calculations of brittle fracture in elastomers, there are still several limitations. Recent experiments in \cite{slootman2020quantifying, slootman2022molecular} have showcased aspects of the damage cascade in elastomers that are previously unclear, namely, that the fracture process zone in some cases cannot be considered negligible. In the phase-field methods, the diffuse nature of fracture is just an approximation. However, from these recent experiments, it is observed that cracks can be physically diffuse, namely, the material constituents deteriorate spatially with a decaying trend as one moves away from the crack plane in the direction of the normal. Thus, we are motivated to explore approaches for the fracture of elastomers where a length scale enters not to regularize the problem, but due to physical considerations.

The GED method has been widely utilized for damage problems, but not as widely for fracture. Some exceptions are \cite{lorentz2017nonlocal, talamini2018progressive}, which will be further discussed in this section. 
Our methodology adopts a strain-based GED modeling framework that has been utilized for linear elastic materials but recasts it in terms of an equivalent polymer chain stretch. We then connect this to local and nonlocal kinematic considerations. In so doing, we utilize components of the constitutive modeling framework discussed earlier in the context of near-incompressibility. We first present a concise overview of strain-based GED and then the extension to the elastomer GED fracture model.

\subsubsection{Strain-based GED models: a summary}\label{GED_strain}
In nonlocal damage models, the process often begins by defining an equivalent strain $\Tilde{\epsilon}$ as a scalar-valued function derived from the infinitesimal strain tensor \cite{de2016gradient}. Subsequently, the nonlocal (equivalent) strain $\Bar{\epsilon}$ is introduced through a spatial averaging procedure over a local neighborhood using the following equation:
\begin{equation}\label{epsilon-bar}
\Bar{\epsilon}(\textbf{x})=\frac{1}{\Psi (\textbf{x})}\int_{\Omega} \Psi(\textbf{y}, \textbf{x})\Tilde{\epsilon} d\Omega.
\end{equation}
Here, $\Psi(\textbf{y}, \textbf{x})$ represents a weight function, often assumed to be homogeneous and isotropic. Following, the damage field is defined as
\begin{equation}\label{damage-strain-function}
\alpha=f(\Bar{\epsilon}),
\end{equation}
where $f(\Bar{\epsilon})$ is a damage function. If we approximate the nonlocal strain $\Bar{\epsilon}$ in equation \eqref{epsilon-bar} using its Taylor series and then integrate, assuming isotropy and truncating after the second-order term, we arrive at
\begin{equation}\label{strain1}
\Bar{\epsilon} = \Tilde{\epsilon} + g\nabla^2\Tilde{\epsilon},
\end{equation}
where $\nabla\Tilde{\epsilon}$ is the spatial gradient of $\Tilde{\epsilon}$ and g is a gradient parameter of the dimension length squared, meaning that $g^{1/2}$ can be considered the characteristic length in the model \cite{peerlings2001critical}.
To avoid high-order continuity requirements for the subsequent  finite element implementation, it is preferable to express equation \eqref{strain1} in the following format
\begin{equation}\label{strain2}
\Bar{\epsilon} - g\nabla^2\Bar{\epsilon} = \Tilde{\epsilon}.
\end{equation}
Solving this equation and the corresponding mechanical equilibrium equation along with the appropriate boundary conditions allows for $\Bar{\epsilon}$ to be determined. This then permits the damage field to be solved using the damage function from equation \eqref{damage-strain-function}. 
\subsubsection{Stretch-based GED for elastomers}
The first aim is to identify a physical quantity that can enable a meaningful and representative construction of a GED model for elastomers. For polymer chains of an elastomeric material, the chain stretch is defined as $\lambda_{ch}={r}/{r_0}$ where $r_0$ is the equilibrium length of a chain and $r$ is the final length of the chain upon deformation. Following the macroscopic-to-microscopic homogenization approach of the 8-chain model introduced by Arruda and Boyce \cite{Arruda-Boyce1993}, we can obtain a chain stretch that represents an average stretch for the polymer chains at every material point in an elastomer (assuming a uniformity in the number of segments that compose each chain)
\begin{equation}\label{Lambda_chain_C}
\lambda_{ch}=\sqrt{\frac{I_1}{3}}.
\end{equation}
Do note that even though here we do not utilize $\lambda_{ch}$ to define a strain energy density, we will utilize it to construct the damage formulation. One could directly extend the suggested approach to strain energies that are also based on the chain stretch.

At this point, we now assume that the polymer chain network (unlike an ordered crystal) is imperfect in its microscopic topology. To account for this network imperfection, we consider a nonlocal approximation to the chain stretch $\lambda_{ch}$, and thus introduce a nonlocal (chain) stretch $\Bar{\lambda}$. Following the process outlined in Section \eqref{GED_strain} in the reference configuration, the length scale $l$ accounts for the size of the region over which the averaging procedure has to take place. 
We can now reformulate \eqref{strain2} in terms of the chain stretch and the nonlocal stretch as follows:\begin{equation}\label{chain_strong}
\Bar{\lambda}-\ell^2\nabla^2\Bar{\lambda}-\lambda_{ch}=0 \, .
\end{equation}

Akin to Eq. \eqref{eq_viscosity}, we add the artificial viscosity to Eq. \eqref{chain_strong} and express the three governing equations as follows,

\begin{equation}\label{strong_forms2}
\begin{split}
\nabla\cdot \mathbf{P} - \bm{T} &=0 \quad \;\;\text{in} \quad {\Omega}_0,\\
- b(\alpha)(J-1) -\frac{p}{\kappa}&=0 \quad \;\;\text{in} \quad {\Omega}_0,\\
\eta\Dot{\Bar{\lambda}} + \Bar{\lambda}-\ell^2\nabla^2\Bar{\lambda}-\lambda_{ch} &=0 \quad\;\; \text{in} \quad {\Omega}_0,\\
\end{split}
\end{equation}
along with the following boundary conditions
\begin{equation}\label{BC_phase_field}
\begin{split}
\bm{u} &= \Bar{\bm{u}} \quad \text{on} \quad \partial_{D}{\Omega}_0,\\
{\bm{T}} &= \Bar{\bm{T}} \,\quad \text{on} \quad \partial_{N}{\Omega}_0,\\
\Bar{\lambda} &= \Bar{\Bar{\lambda}} \quad \text{on} \quad \partial_{D}{\Omega}_0,\\
\bm{n}_0.\nabla\Bar{\lambda} &= 0 \,\quad \text{on} \quad \partial_{N}{\Omega}_0,\\
\end{split}
\end{equation}
Note that equations $\eqref{strong_forms2}_1$ and $\eqref{strong_forms2}_2$ are exactly the same as $\eqref{strong_forms_phase_field}_1$ and $\eqref{strong_forms_phase_field}_2$, respectively.

Utilizing the test functions $v$, $q$, and $\beta$, which are respectively defined in the function spaces $\mathbb{U}$, $\mathbb{P}$, and $\mathbb{L}$ according to \citep{pham2010gradient, brink1996some,le1994numerical}, we obtain the following weak forms

\begin{equation}\label{weak_form_GED}
\begin{split}
\int_{\Omega_0} \mathbf{P}  : \nabla\bm{v} \,dV -\int_{\partial_N\Omega_0}{\bm{T}}_0\cdot\bm{v} \,dA &=0,\\
\int_{\Omega_0}  \left(b(\alpha)\left(J-1\right) -\frac{p}{\kappa}\right)q \,dV &=0,\\ 
\int_{\Omega} \eta\Dot{\Bar{\lambda}}\beta \,dV +\int_{\Omega}\Bar{\lambda}\beta \,dV + \int_{\Omega}\ell^2\nabla\Bar{\lambda}\cdot\nabla\beta \,dV - \int_{\Omega}\lambda_{ch}\beta \,dV &=0.
\end{split}
\end{equation}
Again, equations $\eqref{weak_form_GED}_1$ and $\eqref{weak_form_GED}_2$ are the same as $\eqref{weak_form_phase_field}_1$ and $\eqref{weak_form_phase_field}_2$, respectively.

\subsubsection{Damage function}
A significant departure from the phase-field method is that the stretch-based GED necessitates the calculation of the nonlocal stretch using Eq. \eqref{chain_strong}. Then, the damage field $\alpha$ is subsequently evaluated using a pre-defined damage function. Various damage functions have been proposed in the literature \cite{peerlings1996gradient, verhoosel2011isogeometric}; a commonly utilized one \cite{sarkar2019comparative} is adapted here as follows:

\begin{equation}\label{damage_function}
\alpha(\Bar{\lambda}) = 
\begin{cases} 
0 & \text{if } \Bar{\lambda} < \lambda_{cr} \\
1 - \frac{\lambda_{cr}-1}{\Bar{\lambda}-1}\left(1 - c + c e^{-\gamma(\Bar{\lambda} - \lambda_{cr})}\right) & \text{if } \Bar{\lambda} \geq \lambda_{cr} 
\end{cases},
\end{equation}
where $\lambda_{cr}$ can be considered as the critical stretch marking the onset of the damage \cite{Arruda-Boyce1993}. 

Initial studies for polymer chain statistics determined a critical chain stretch by only considering entropic contributions. But more recently, enthalpic contributions have also been introduced in such considerations, which indicate further stretching of the chain and subsequent damage \cite{li2020variational,mao2018theory,mulderrig2023statistical}. Fully mapping the details of the aforementioned work in statistical mechanics towards a continuum damage model will be the focus of a subsequent work.  %

\section{Numerical implementation}\label{Section:NumericalImplementation}
Our numerical implementation of the phase-field fracture formulation, gradient-enhanced damage modeling, and J-integral calculation are built upon the \texttt{FEniCS} finite element platform \cite{alnaes2015fenics}, leveraging the automatic functional differentiation tool provided by the Unified Form Language (UFL). The Python scripts developed for this purpose are available upon request.

\subsection{Interpolation and solution approach}\label{Subsection:discretization}
Both the phase-field formulation and the stretch-based GED formulation are implemented in \texttt{FEniCS} in a mixed finite element model. Thus, there are specific choices that we need to consider regarding the interpolation scheme, as both formulations consider the limit of near-incompressibility.  For the phase-field model, following \cite{li2020variational}, the displacement, pressure, and damage fields are approximated as

\begin{equation}\label{eq_displacement}
\bm{u}(\bm{x})= \sum_{i=1}^{N} N_i(\bm{x}) \bm{u}_i, \quad\quad\quad\quad
p(\bm{x})= \sum_{i=1}^{N} N_i(\bm{x}) p_i, \quad\quad\quad\quad
\alpha(\bm{x})= \sum_{i=1}^{N} N_i(\bm{x}) \alpha_i,
\end{equation}
where $N_i(\bm{x})$ are the shape functions corresponding to node $i$, and $\bm{u}_i$, $p_i$, and $\alpha_i$ denote the nodal values corresponding to the displacement, pressure, and damage fields, respectively. Note that in the gradient-enhanced damage model, instead of the damage variable being a primary nodal unknown, the nonlocal stretch is a nodal unknown $\Bar{\lambda}(\bm{x})= \sum_{n=1}^{N} N_i(\bm{x}) \Bar{\lambda}_i$. Here, $\Bar{\lambda}_i$ are the nodal values corresponding to the nonlocal stretch field. To resolve issues regarding spurious oscillations that might arise due to the inf-sup condition, we utilize a Taylor-Hood space (not of the lower order), where we employ quadratic interpolation for the displacement field and linear interpolation for the pressure, damage (for the case of phase-field) and nonlocal stretch (for the case of GED) fields.

As we assume loading under quasi-static conditions, we employ a staggered scheme for the solution approach, that loops over a load-ramping function. The staggered scheme consists of an exterior loop that checks for convergence between successive iterations, and it includes two interior successive loops for the solution of the staggered problem as explained below. In the first interior loop, we solve the mechanical equilibrium and Lagrange multiplier equations coupled together using the \texttt{FEniCS} built-in nonlinear non-constrained solver (\texttt{SNES} with method: \texttt{newtontr}), assuming a fixed value for $\alpha$ (in the phase-field) or $\Bar{\lambda}$ (in GED). We respectively denote these fixed values as $\alpha_{j-1}$ and $\Bar{\lambda}_{j-1}$, with $j$ being the counter for the exterior loop. We use the critical point (\texttt{cp}) line search in the \texttt{SNES} solver, and \texttt{absolute}, \texttt{relative}, and \texttt{solution} tolerances are all set to $10^{-10}$ with a \texttt{maximum iterations} of $300$. In the subsequent internal loop, we solve the damage equation (in the phase-field) or the nonlocal stretch equation (in GED) using the \texttt{FEniCS} built-in nonlinear constrained solver (\texttt{SNES} with method: \texttt{vinewtonssls} with \texttt{absolute} and \texttt{relative} tolerances equal to $10^{-10}$ and \texttt{maximum iterations} of $300$) this time keeping the displacement and pressure fields fixed at $\mathbf{u}_{j}$ and $p_j$. Then, in the external loop for the staggered scheme, we check if the sup norm error $|\alpha_{j} - \alpha_{j-1}|_{\infty}$ (in the phase-field) or $|\Bar{\lambda}_{j}-\Bar{\lambda}_{j-1}|_{\infty}$ (in GED) is less than $10^{-3}$, which serves as the third convergence criterion. If this condition is satisfied, the code advances to the next loading time step, and the $\mathbf{u}$, $p$, and $\alpha$ or $\Bar{\lambda}$ pass to the next step to serve as the initial fixed values for solving the internal loops. The maximum number of iterations for this criterion is also 300.

\subsection{Domain J-integral implementation}\label{Subsection:discretization}

As the scope of this paper is to verify the phase-field and GED approaches for fracture in elastomeric materials under different loading and parameter choices, we implement a domain J-integral \cite{li1985comparison, bouklas2015effect} to be utilized as part of solution post-processing. Our choice of the domain J-integral implementation is due to the fact that it considers area integrals instead of path integrals in two dimensions (it considers volume integrals instead of surface integrals in three dimensions), as would be the case for the traditional J-integral approach \cite{rice1968path}. 

\begin{figure}[h!]
    \centering
    \includegraphics[width=0.4\linewidth]{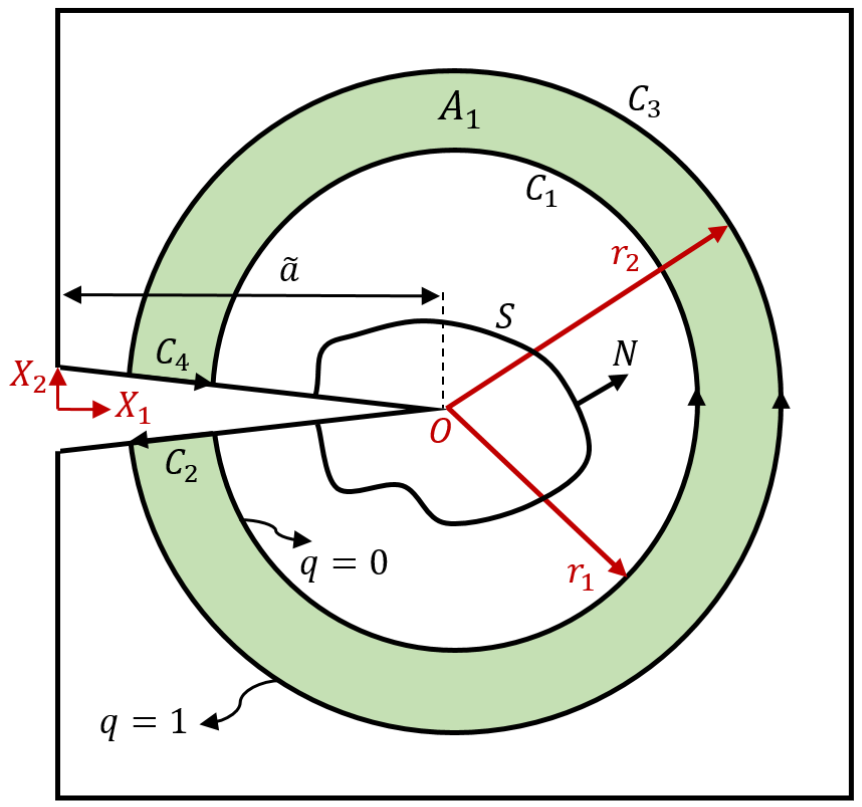}
    \caption{ A schematic for the domain J-integral in a sharp crack setting in the reference configuration.}
    \label{fig:J-integral}
\end{figure}

Utilizing the deteriorated strain energy density function corresponding to the phase-field or GED approaches, which we will generically denote as $\widetilde{\mathcal{W}}$, the J-integral for an arbitrary fracture, as illustrated in Fig. \ref{fig:J-integral}, is given in the reference configuration as \cite{rice1968path, rice1968mathematical}
 \begin{equation}\label{J_integral_path}
J = -\frac{d\psi}{d\Tilde{a}}=\int_{S}  (\widetilde{\mathcal{W}} N_1-P_{iJ}N_J\frac{\partial x_i}{\partial X_1}) \,d S.
\end{equation}
Here, $\Tilde{a}$ is the crack length, $S$ is a path around the tip as depicted in Fig. \ref{fig:J-integral}, $N$ is the unit normal vector to this path, and $P_{iJ}$ is the first Piola-Kirchhoff stress tensor. This can be adapted to the domain J-integral formulation as
 \begin{equation}\label{J_integral}
J =\int_{A_1}  (-\widetilde{\mathcal{W}} \frac{\partial q}{\partial X_1}+P_{iJ}F_{i1}\frac{\partial q}{\partial X_J}) \,d A,
\end{equation}
where $A_1$ is a closed area around the tip. For simplicity, we assume this region is an annular area between $r_1$ and $r_2$. Moreover, $q$ is a sufficiently smooth function in $A_1$, ranging from zero
at $C_1$ to unity at $C_3$, as shown in Fig. \ref{fig:J-integral}. The derivation details for Eq. \eqref{J_integral} can be found in Bouklas \textit{et al.} \cite{bouklas2015effect} (2015). This so-called 
domain J-integral can be used to calculate the energy release rate during crack growth under Mode-I loading, and for a sharp crack model (as depicted in Fig. \ref{fig:J-integral}). We note that as the damage region should be contained in the interior of $C_1$, we can, without loss of generality, substitute $\widetilde{\mathcal{W}}$ with $\widetilde{\widetilde{\mathcal{W}}}$ since the material within region $A_1$ has not accumulated any damage. The domain J-integral approach has previously been utilized by the authors in a multiphysics setting for large deformation poroelasticity in elastomers \cite{bouklas2015effect}.

\section{Results and Discussion}\label{Section:results}
In finite element simulations, all normalized parameters are represented by $\widehat{\left(\bullet\right)}$. The normalized first P-K stress and hydrostatic pressure are expressed as:
\begin{equation}
\begin{split}
    \mathbf{\widehat{P}} = \frac{\mathbf{P}}{\mu}, \quad \widehat{p} = \frac{p}{\mu} \quad . 
\end{split}
\end{equation}
Additionally, we can define the characteristic length of the domain as $L_0$ and consider this as a reference length.

\begin{figure}[h!]
    \centering
    \includegraphics[width=0.65\linewidth]{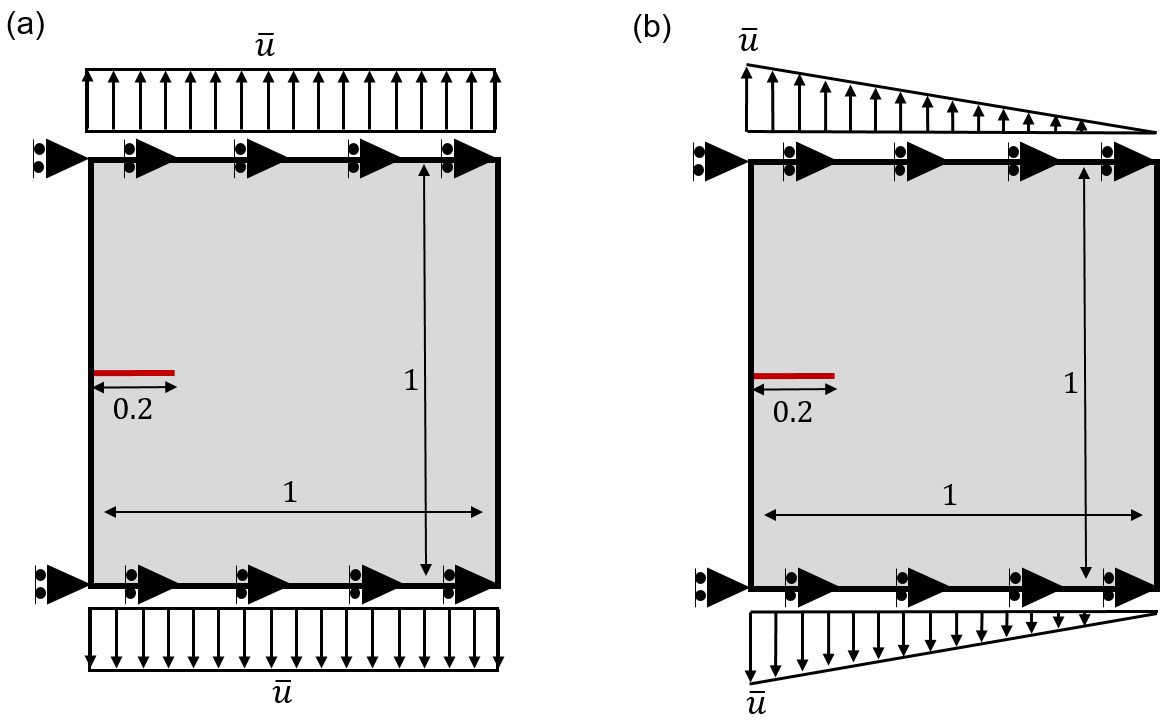}
    \caption{A rectangular domain with an edge-crack in plane strain conditions subject to (a) uniform loading, and (b) triangular-type loading.}
    \label{fig:loading}
\end{figure}

In our numerical examples, we analyze an $L_0\times L_0$ square domain in plane strain conditions, where $L_0=1$, subjected to two types of displacement-controlled loading: a uniform tensile load, and a triangular-type load, as illustrated in Fig. \ref{fig:loading}. In both cases, we consider the coordinate origin at the middle of the left boundary, as shown in Fig. \ref{fig:J-integral}. For the uniform loading, we have $\Bar{u}_1=0$ and $\Bar{u}_2=0.35$ at $X_2=0.5$, and $\Bar{u}_1=0$ and $\Bar{u}_2=-0.35$ at $X_2=-0.5$. In addition, $\Bar{T}=0$ at $X_1=0$ and $X_1=1$. For the triangular-type loading, we have $\Bar{u}_1=0$ and $\Bar{u}_2=0.35(1-X_1)$ at $X_2=0.5$, and $\Bar{u}_1=0$ and $\Bar{u}_2=-0.35(1-X_1)$ at $X_2=-0.5$. Similar to uniform loading, $\Bar{T}=0$ at $X_1=0$ and $X_1=1$. We chose these loading conditions to investigate the problem under contrasting circumstances, as uniform loading tends to destabilize the problem, whereas triangular-type loading provides a more stable environment for crack growth.

For the phase-field modeling examples, the base case scenario includes an internal length scale of $\ell=0.04$, a fracture energy of $G_c/\mu=1$ for 2D simulations, and a maintained ratio of $\kappa/\mu = 1 \times 10^3$, which corresponds to $\nu = 0.4995$ to ensure near-incompressibility. For GED modeling, we use the same values for the length scale, bulk modulus, and shear modulus.

The first four subsections that follow involve numerical experiments utilizing the phase-field method, and the subsequent last subsection involves numerical experiments utilizing the GED approach. We commence with an unstable crack propagation example, followed by a stable propagation validation study utilizing the domain J-integral. Subsequently, we delve into exploring the impact of artificial viscosity on the critical energy release rate during propagation. Furthermore, we explore the discrepancies between diffuse and discrete approximations of the initial fracture representation. Lastly, we model crack propagation using the GED approach and analyze the effect of artificial viscosity on the critical energy release rate as well as other issues that we encounter with this type of simulation.

\subsection{Viscous dissipation in phase-field calculations}\label{subsection:example1}

On top of the development of staggered schemes to enable the solution of the mixed formulation, there are several other numerical challenges that need to be addressed for the phase-field approach to brittle fracture modeling. Importantly, under a quasistatic setting, it is very common to encounter unstable crack growth. Typically, when utilizing a Newton-type solver for a nonlinear problem, we aim for our initial guess for every time step to be as accurate as possible. Clearly, this is hard to achieve during unstable crack growth, leading to convergence issues in the nonlinear solution scheme. In this subsection, we will explore some problems that exhibit unstable crack growth and some problems that exhibit stable crack growth.

\begin{figure}
    \centering
    \includegraphics[width=0.7\linewidth]{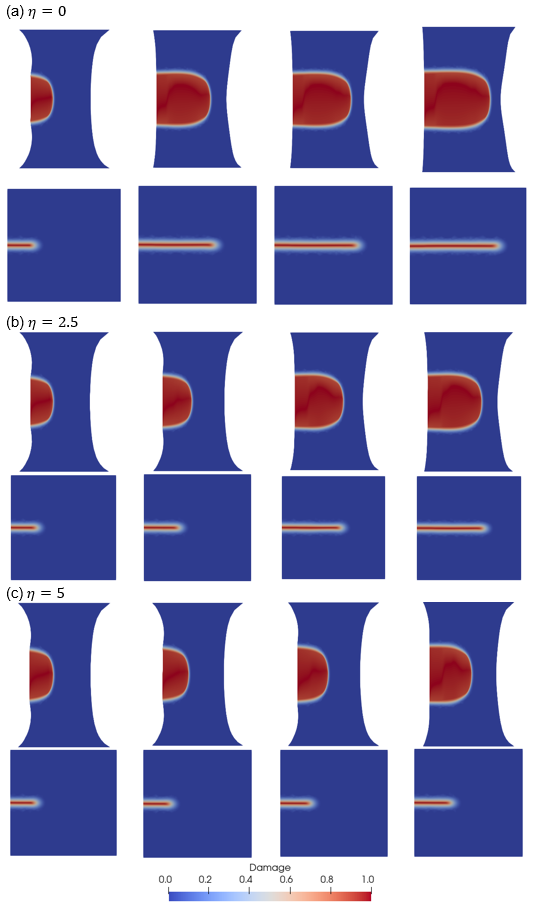}
    \caption{The result of the phase-field simulation for the edge-crack sample under uniform loading in the current (top rows) and reference (bottom rows) configurations for viscosity parameter with a value of (a) 0, (b) 2.5, and (c) 5. Each row shows the first four time steps after the onset of the crack propagation. Note that at the initial state, the crack is represented diffusely.}
    \label{fig:example1-contours}
\end{figure}

Our investigation commences featuring a geometry and loading corresponding to Fig. \ref{fig:loading} (a), with material properties and boundary conditions corresponding to the base case. We place a pre-existing diffuse crack at $0\leq X_1 \leq0.2$ and $X_2=0.5$ by setting $\alpha=1$ as the lower bound of our bounded \texttt{SNES} (method: \texttt{vinewtonssls}) nonlinear solver. We explore a range of artificial viscosities as $\eta=\left\lbrace0, 1, 2.5, 5\right\rbrace$, and Table \ref{tab:table1} enumerates the crack length progression across consecutive steps following the onset of crack growth for these four cases. Fig \ref{fig:example1-contours} illustrates the initial four steps of crack propagation for the cases corresponding to $\eta=\left\lbrace0, 2.5, 5\right\rbrace$. Yellow cells within the table highlight instances wherein significant convergence challenges were encountered. Specifically, as discussed in Section \ref{Subsection:discretization}, we have set three distinct convergence criteria in the staggered scheme, and the yellow cells indicate our inability to meet the third criterion $|\alpha_{j} - \alpha_{j-1}|_{\infty}<10^{-3}$ within 300 iterations (maximum number of iterations) in that step. The onset of this phenomenon indicates unstable crack growth. 

\begin{table}[htbp]
  \caption{Consecutive steps of crack propagation from the onset of crack growth to the fully broken state for four simulations with distinct artificial viscosity parameters. Yellow cells highlight instances wherein significant convergence challenges were encountered.}
  \label{tab:table1}
  \centering
  \includegraphics[width=0.9\textwidth]{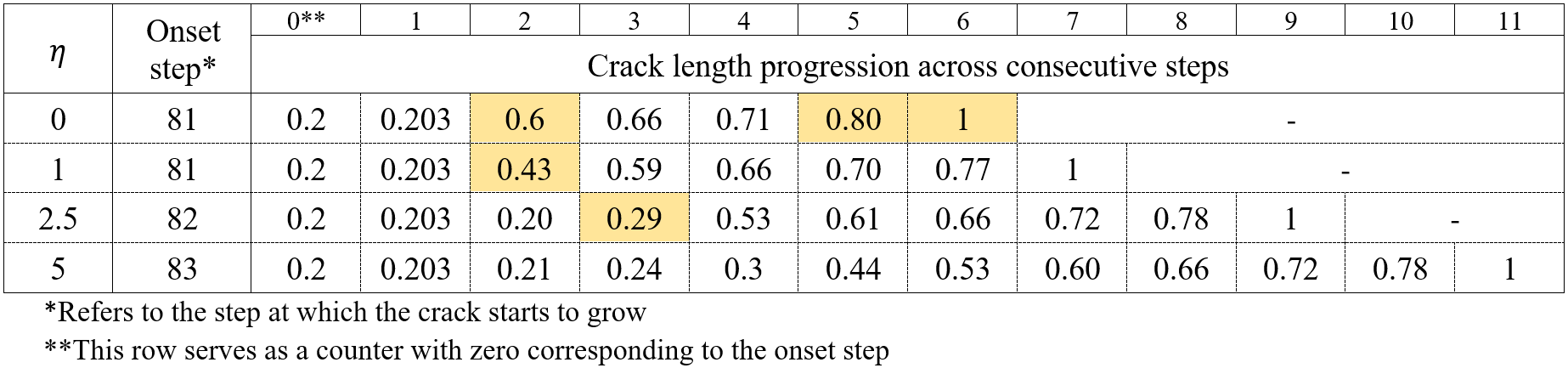}
\end{table}

\begin{figure}
    \centering
    \includegraphics[width=0.5\linewidth]{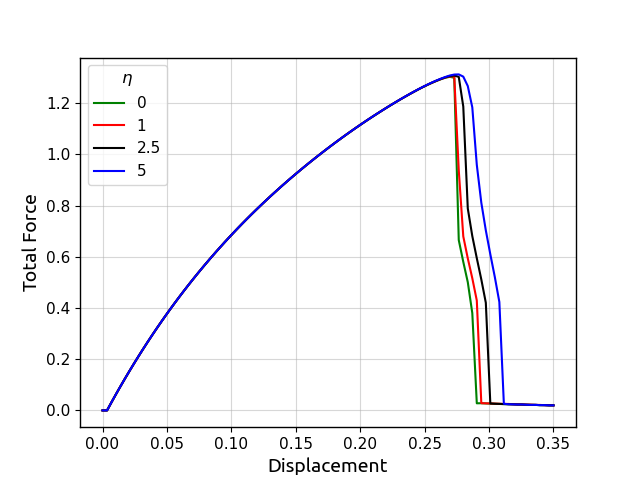}
    \caption{Load-displacement curves for four simulations with different artificial viscosity parameters (range from 0 to 5) for the rectangular domain with an edge-crack under uniform loading. The edge-crack in the initial state is represented diffusely.}
    \label{fig:example1-graph}
\end{figure}

To mitigate this issue in the simulation, one approach is the incorporation of artificial viscosity \cite{miehe2010rate} as per Eq. \ref{eq_viscosity}, which serves to penalize the rapid growth of the damage variable pointwise. As the artificial viscosity parameter $\eta$ increases from $0$ to $5$ according to Table \ref{tab:table1}, the number of crack propagation steps from the onset of the crack propagation until ultimate failure also increases (from 6 steps when $\eta=0$ to 11 steps when $\eta=5$, as counted in Table \ref{tab:table1}). This increase in the number of steps signifies a decline in crack tip velocity, indicating a stabilization of crack growth. Note that this solution is meant to enable the acquisition of converged results, even though there is no clear physical backing for the introduction of artificial viscosity.
Analyzing the onset column of Table \ref{tab:table1}, we observe that as the artificial viscosity parameter increases, the crack propagation onset is slightly delayed. This delayed onset of crack propagation is clearly illustrated in Fig. \ref{fig:example1-graph}, where the x-axis represents the vertical displacement of the top left corner of the body, and the y-axis represents the total force exerted on the upper boundary. Note that the units are normalized.

The shift in crack propagation onset implies that we are effectively altering the fracture toughness of the material, a key material property. In the subsequent sections, we will meticulously examine this effect. However, prior to this analysis, it is crucial to first evaluate our methodology for calculating the J-integral.

\subsection{Fracture energy verification for the case of no viscous contributions}\label{subsection:example2}
\begin{figure}
    \centering
    \includegraphics[width=0.75\linewidth]{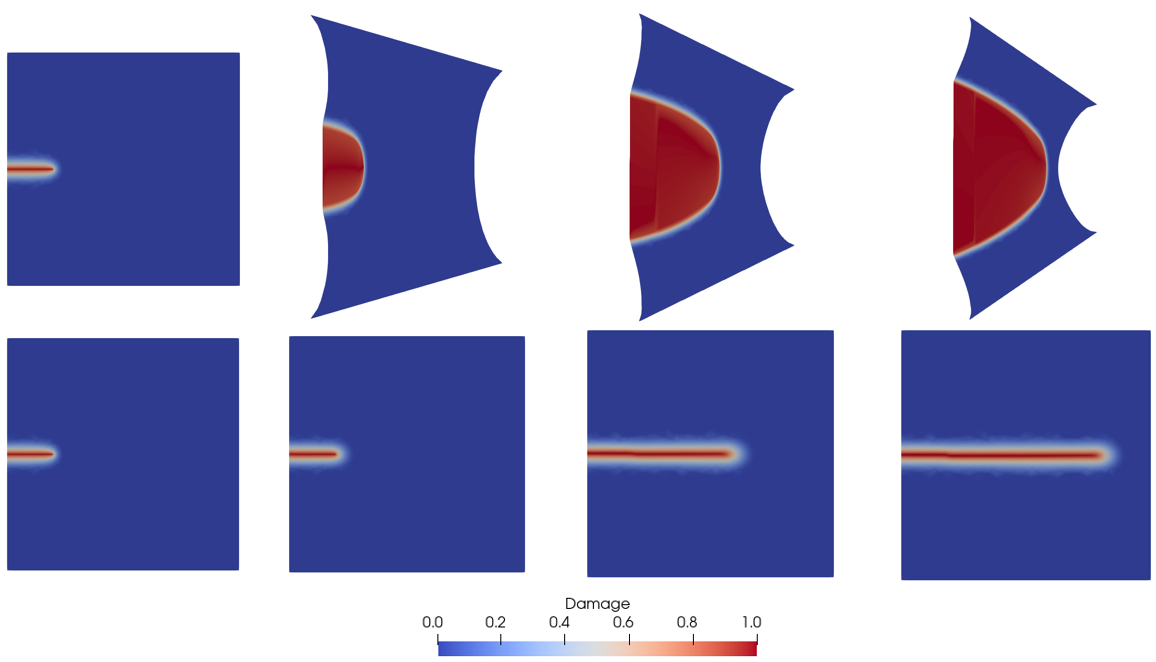}
    \caption{The result of the phase-field simulation for the edge-crack sample under triangular-type loading with no artificial viscosity at steps 1, 30, 50, and 70 from left to right. Note that at the initial state, the crack is represented diffusely.}
    \label{fig:example2-contours}
\end{figure}

\begin{figure}
    \centering
    \includegraphics[width=0.5\linewidth]{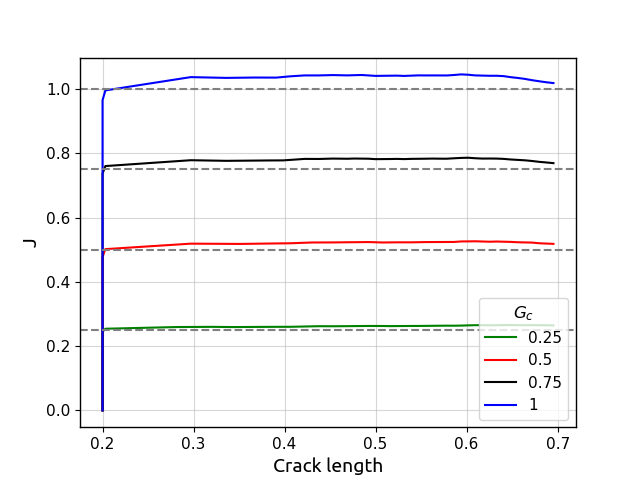}
    \caption{For an edge-crack sample with triangular-type loading, domain J-integral evaluations with respect to measured crack length for different values of imposed critical energy release rate ($G_c$ range from 0.25 to 1). The dashed lines show the imposed $G_c$ values.}
    \label{fig:example2-graph1}
\end{figure}

In this section, our objective is to verify the methodology we employ for calculating the J-integral. To conduct this validation without the influence of artificial viscosity, we employ the triangular-type loading configuration shown in Fig. \ref{fig:loading}(b). This specific loading condition is selected to ensure the solver convergence. The result for the base case is depicted in Fig. \ref{fig:example2-contours}.
We anticipate that as the crack starts to grow, the energy release rate (J-integral) will be a constant value that is equal to $G_c$, which is an input to the model. To calculate the J-integral, we position the centers of the circular paths at the center of the domain as depicted in Fig. \ref{fig:J-integral}, and set $r_1$ and $r_2$ to be $0.4$ and $0.47$, respectively. Reviewing Fig. \ref{fig:example2-graph1}, we observe that as the crack initiates and propagates, the J-integral indeed approaches a constant value, matching $G_c$ for each case (four cases were tested: $G_c=\left\lbrace 0.25, 0.5, 0.75, 1\right\rbrace$). This outcome confirms the accuracy and functionality of our J-integral calculation method. It is important to note that the calculation of the J-integral is valid only when the crack is confined within the area enclosed by path $C_1$ in Fig. \ref{fig:J-integral}. Once the crack starts to enter the area $A_1$, the results are no longer meaningful.

\begin{figure}
    \centering
    \includegraphics[width=0.6\linewidth]{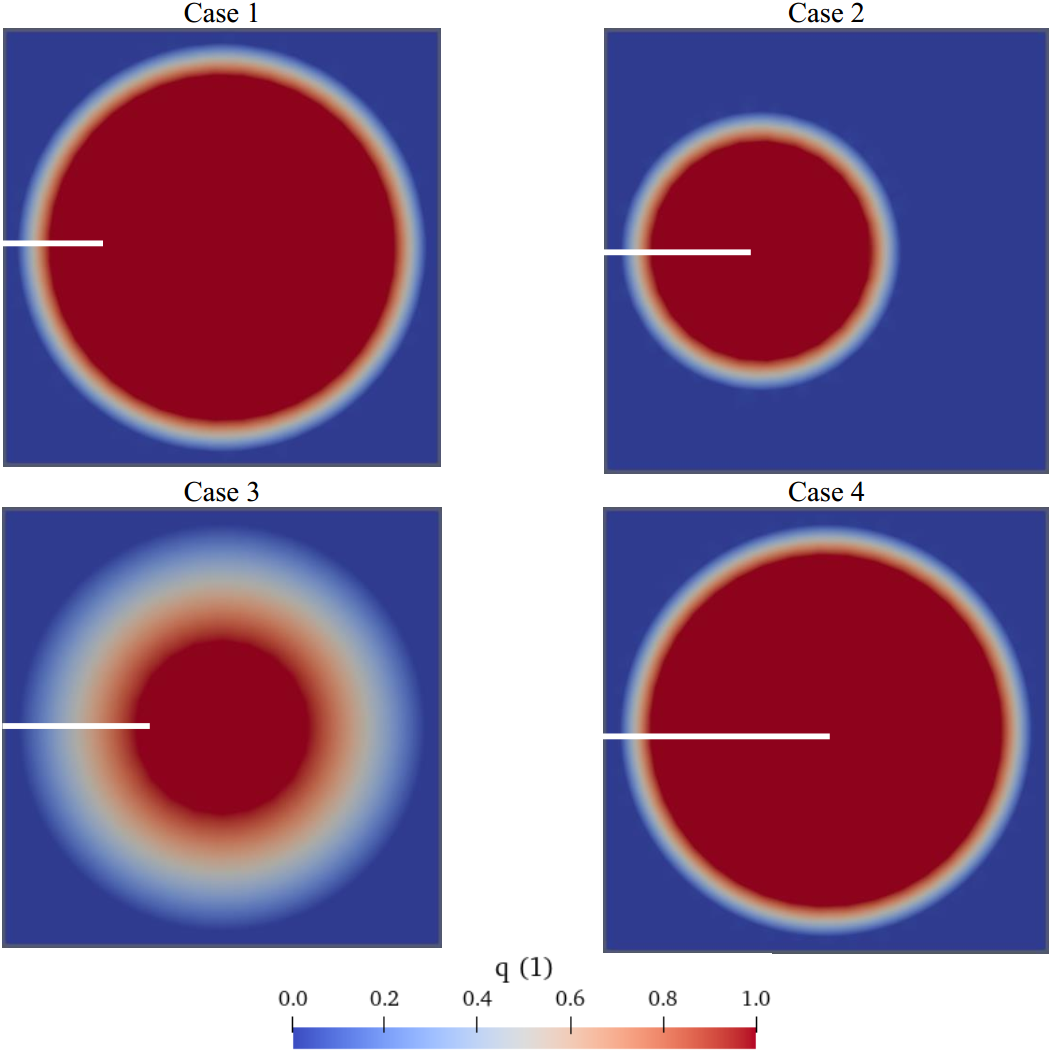}
    \caption{Plots of the $q$ functions utilized in the domain J-integral for four different cases with crack length, circle center location, and circle radii as 1)$\Tilde{a}=0.2$, $O=0.5$, $r_1=0.4$, $r_2=0.47$ 2) $\Tilde{a}=0.35$, $O=0.35$, $r_1=0.25$, $r_2=0.32$ 3) $\Tilde{a}=0.35$, $O=0.35$, $r_1=0.2$, $r_2=0.47$ 4) $\Tilde{a}=0.5$, $O=0.5$, $r_1=0.4$, $r_2=0.47$.}
    \label{fig:example2-four-scenario}
\end{figure}

\begin{figure}
    \centering
    \includegraphics[width=0.5\linewidth]{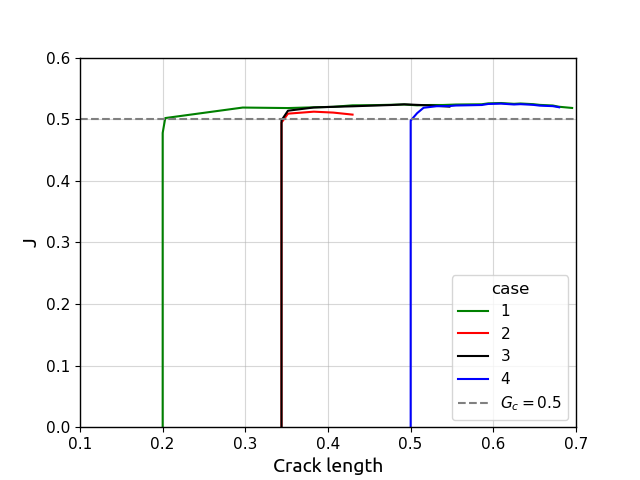}
    \caption{Evaluation of the  domain J-integral during crack propagation for an edge-crack sample under triangular-type loading with $G_c=0.5$ corresponding to the cases shown in Fig. \ref{fig:example2-four-scenario}.}
    \label{fig:example2-graph2}
\end{figure}

Furthermore, we seek to verify that our J-integral implementation leads to path-independent results. To assess this, we explore four distinct scenarios where the center of the circles, the thickness of $A_1$, and the crack tip location vary, as depicted in Fig. \ref{fig:example2-four-scenario}. 
The J-integral curves for these four scenarios are shown in Fig. \ref{fig:example2-graph2}. In each case, as the crack propagation initiates, the J-integral evaluation remains in near proximity to $G_c$, until the crack intersects with the J-integral domain area. The discrepancy of the predicted value at steady-state is approximately $4.5\%$ which is in line with our expectations following Ang \textit{et al.} (2022) \cite{ang2022stabilized}.

\subsection{Effect of viscous contributions on fracture energy predictions}\label{subsection:example3}

\begin{figure}[h!]
    \centering
    \includegraphics[width=0.87\linewidth]{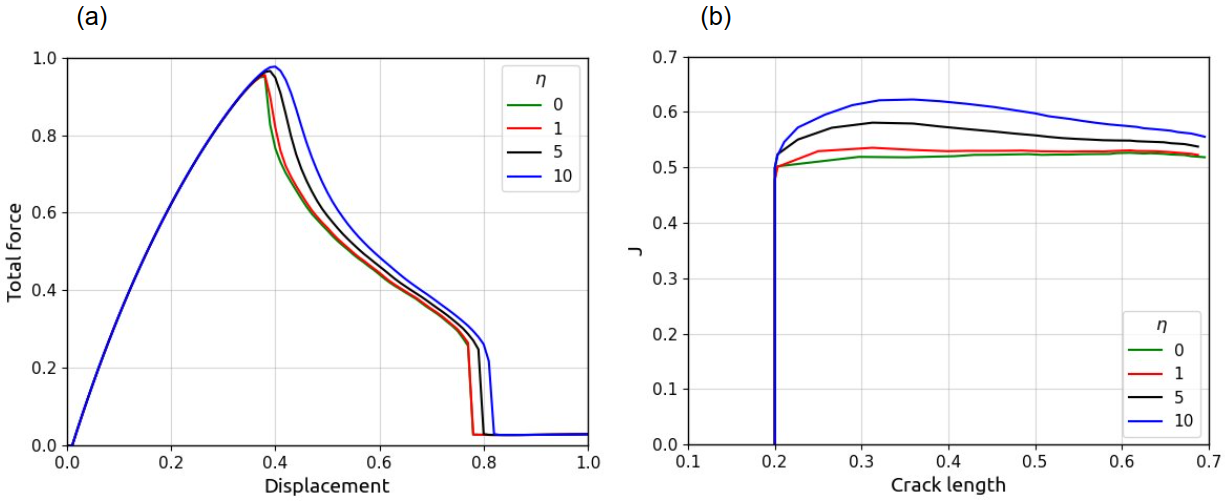}
    \caption{The effect of artificial viscosity parameter on (a) force-displacement curve and (b) energy release rate as obtained from the domain J-integral.  The sample is an edge-crack sample with a diffuse representation of the crack at the initial state under triangular-type loading.}
    \label{fig:example3-graph}
\end{figure}

We noted at the end of Section \ref{subsection:example1} that introducing artificial viscosity alters the energy release rate of the problem. To examine this impact, we configure the problem as shown in Fig. \ref{fig:loading}(b) using triangular-type loading. We conduct a series of simulations using the base case material properties with varying artificial viscosity parameters. Fig. \ref{fig:example3-graph}(a) illustrates the total force on the top boundary versus the vertical displacement of the top left corner. Based on this result, we can infer that introducing artificial viscosity to the material increases its resistance to fracture. For instance, at a displacement of $0.5$, the total force is approximately $0.55$ for $\eta=0$, whereas it rises to $0.65$ for $\eta=10$.

Fig. \ref{fig:example3-graph}(b) illustrates the energy release rate across different artificial viscosity parameters. The plot highlights two important findings. First, adding artificial viscosity results in an increased energy release rate during crack propagation: for instance, with $\eta=10$, the peak energy release rate rises by nearly $20\%$, a considerable increase. Second, the energy release rate is no longer constant when artificial viscosity is included, showcasing its dynamic behavior. These insights emphasize the significant influence of artificial viscosity on the prediction of critical energy release rate through phase-field fracture calculations.

Without artificial viscosity, crack propagation is the sole mode of energy dissipation in the problem. However, with artificial viscosity as another dissipative mechanism, energy absorption occurs, leading to delayed crack initiation and an increase in the material's energy release rate. The variant of the J-integral that we have implemented is not sensitive to other dissipation mechanisms aside from crack propagation. Thus, it lumps the effect of artificial viscosity in the energy release rate predictions.
Furthermore, the influence of artificial viscosity is more pronounced in the initial steps of crack propagation. This is evident in Fig. \ref{fig:example1-contours}(a) where, in the case of no artificial viscosity, there is a significant jump (sharp and huge movement of the crack) between the first and second frames (by first/second frame, we refer to the first/second picture from the left in a figure, and by step, we refer to a specific time step in the numerical simulation). The artificial viscosity, dependent on the difference in the damage field between the current and previous steps, is crucial in moderating this abrupt change. Consequently, for cases (b) and (c) in Fig. \ref{fig:example1-contours} where artificial viscosity is introduced, this sharp jump is mitigated.
On the other hand, in Fig. \ref{fig:example1-contours}(a), there is not a substantial crack propagation evident in frames 3 and 4. Therefore, the impact of artificial viscosity in these frames is relatively minor, as the changes in the damage field are less drastic compared to the change between frames 1 and 2, which are the initial steps of crack propagation.

\subsection{Sensitivity at crack initiation}

In this section, we examine the influence that the pre-crack representation -- either discrete or diffuse -- has on crack initiation. Thus, we replace the diffuse representation of the pre-crack in the domain with a notch having $0.02$ thickness, a tip radius of $0.02$, and a length of $0.2$ from the tip to the left edge of the domain. All other parameters and conditions remain consistent with the previous section. We conduct the simulation once more, this time setting $\eta=5$.

\begin{figure}[h!]
    \centering
    \includegraphics[width=0.8\linewidth]{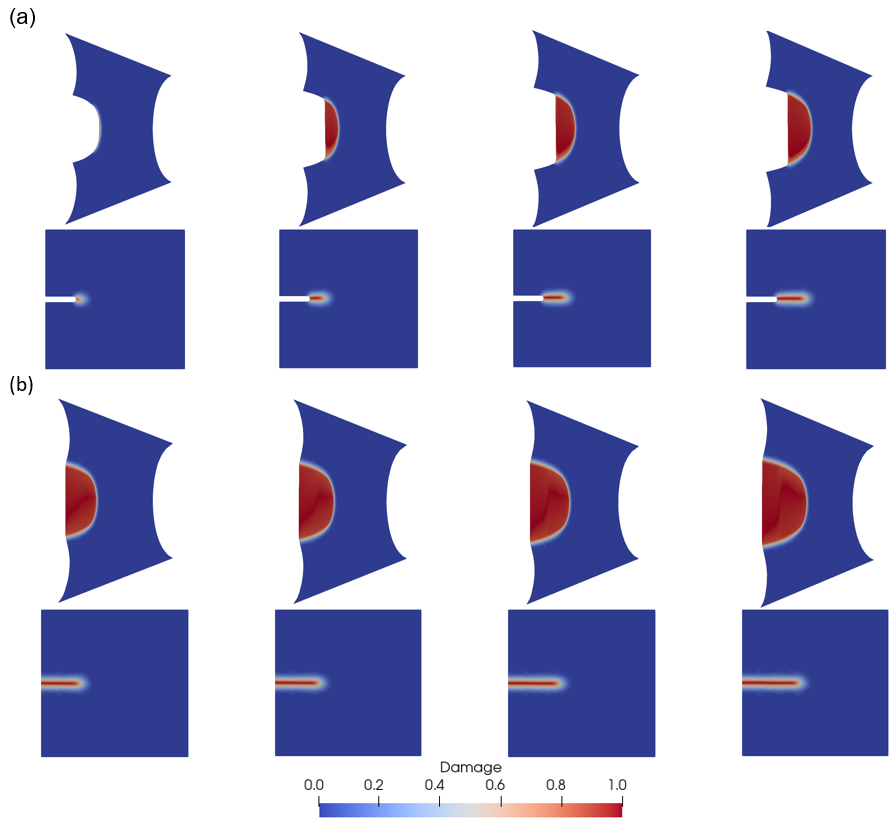}
    \caption{Crack propagation for an edge-crack sample under triangular-type loading for four consecutive time steps after the onset of crack propagation, depicted for (a) discrete and (b) diffuse cracks. The artificial viscosity parameter is set at  $\eta=0.5$.}
    \label{fig:example4-contours}
\end{figure}

\begin{figure}[h!]
    \centering
    \includegraphics[width=0.5\linewidth]{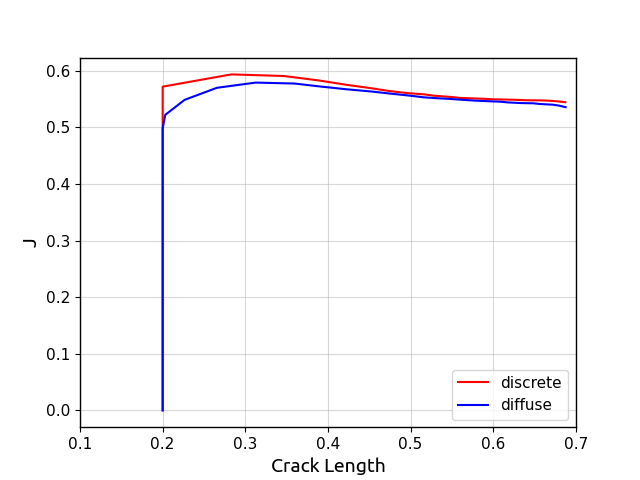}
    \caption{A comparison of the energy release rate during crack propagation for an edge-crack sample under triangular-type loading, where the crack is initially represented in a discrete or diffuse fashion. The artificial viscosity parameter is set at  $\eta=0.5$. }
    \label{fig:example4-graph}
\end{figure}

Fig. \ref{fig:example4-contours} presents the results of two simulations involving discrete and diffuse fractures across four consecutive time steps, both in deformed and undeformed scenarios post-crack initiation. In the discrete pre-crack scenario, crack propagation begins at step $43$, with the crack length increasing from $0.2$ to $0.28$. On the other hand, in the diffuse pre-crack scenario, crack propagation starts at step 41, and the crack length increases from $0.2$ to $0.22$. This suggests that discrete fractures require more energy for crack initiation compared to a diffuse representation in this setting. Fig. \ref{fig:example4-graph} offers a clearer comparison: in the discrete case, crack propagation starts at a higher J-integral value of $J=0.58$, whereas for the diffuse case, crack propagation begins at $J=0.52$. Given that the critical value for crack propagation is $G_c=0.5$, we note a larger difference between J and $G_c$ in the discrete case -- roughly $16\%$. This difference stems from the fact that in discrete cases, energy is not only needed for crack propagation but also for damage nucleation. Consequently, energy dissipates during crack initiation before the crack starts growing. After the initial steps and the crack nucleation in the discrete case, the two curves converge to the same values. 

\subsection{Gradient-enhanced damage calculations}

\begin{figure}[h!]
    \centering
    \includegraphics[width=0.5\linewidth]{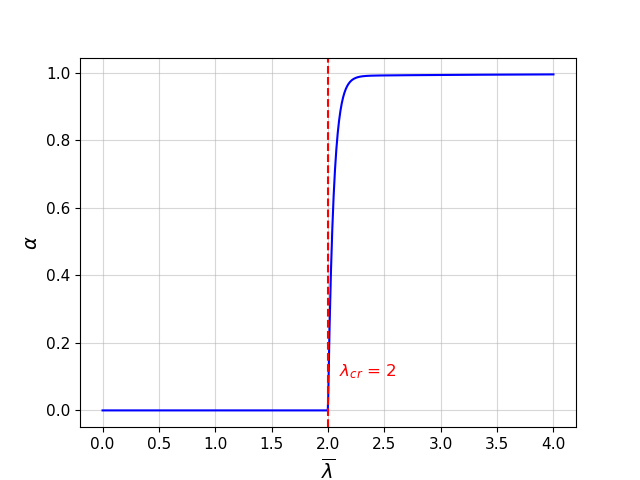}
    \caption{Damage function utilized in the GED framework plotted using the values $c=0.99$ and $\gamma=20$.}
    \label{fig:damage_function}
\end{figure}

We now shift our focus from the phase-field simulation to the GED method for simulating crack propagation in elastomers. Unlike the phase-field method where the   critical energy release rate $G_c$ (which is  a macroscopic quantity) is an input of the problem, the proposed GED formulation aims to recapitulate the fracture process from chain--level damage considerations. Thus, the calculation of the critical energy release rate can occur as an outcome of the solution process. In the GED approach, we prescribe the critical chain stretch $\lambda_{cr}$ in the damage function (Eq. \eqref{damage_function}), which signifies the beginning of network degradation and is a material parameter that can be determined through statistical mechanics; having set $\lambda_{cr}=2$, $c=0.99$, and $\gamma=20$, Fig. \ref{fig:damage_function} illustrates the corresponding damage function with respect to the nonlocal chain stretch.

\begin{figure}
    \centering
    \includegraphics[width=0.7\linewidth]{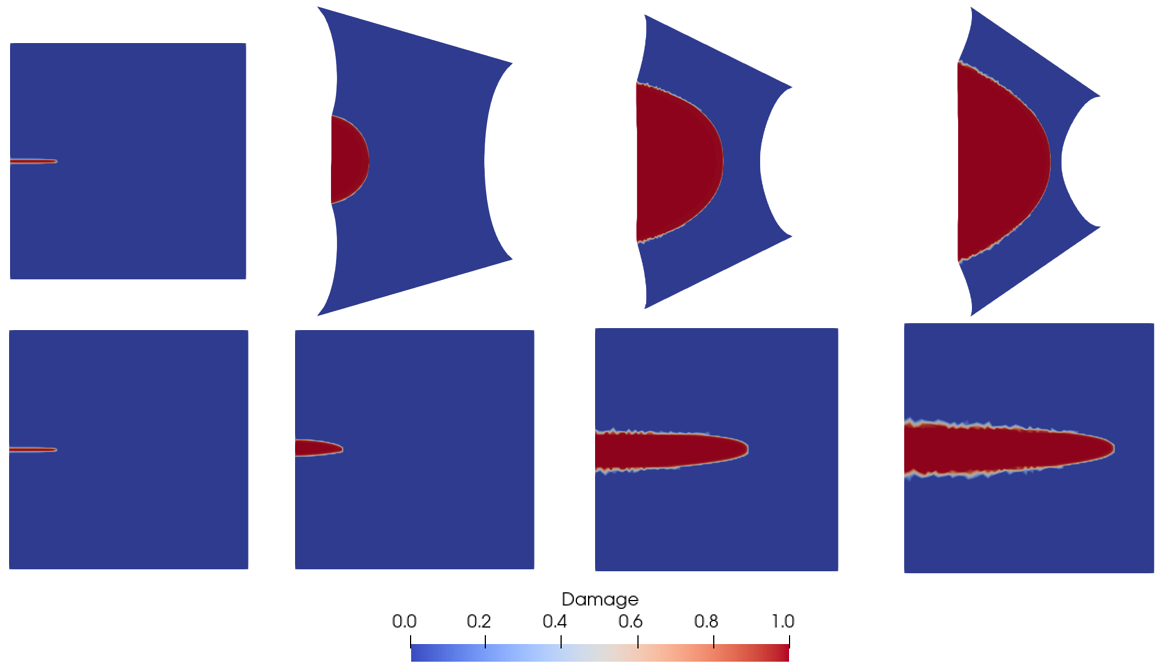}
    \caption{The result of GED simulations for an edge-crack sample with an initially diffuse crack under triangular-type loading and with no artificial viscosity. Snapshots correspond to steps 1, 33, 37, and 75 from left to right.}
    \label{fig:example6-contour}
\end{figure}

For the GED simulation, we use triangular-type loading along with the same material properties that we had in the base case. But now, instead of prescribing the $G_c$, we set the critical chain stretch as $\lambda_{cr}=2$. Additionally, like the diffuse crack in the phase-field modeling, we place a pre-existing diffuse crack at $0\leq X_1 \leq0.2$ and $X_2=0.5$ by setting $\Bar{\lambda}=2.3$, which results in $\alpha=1$ according to Fig. \ref{fig:damage_function}. This initialized $\alpha$ field serves as the lower bound of our bounded \texttt{SNES} (method: \texttt{vinewtonssls}) nonlinear solver. The results of the GED simulation are presented in Fig. \ref{fig:example6-contour}. While the outcomes are largely similar to those of the phase-field simulation, a notable distinction arises: the stretch-based GED, akin to its strain-based counterpart, experiences broadening issues (a non-physical widening of the damage zone in the fully fractured region, normal to the crack propagation plane). This transverse propagation of damage becomes evident in the undeformed configuration. To be more specific, in Fig. \ref{fig:example6-contour} and for the undeformed configuration (second rows), the damage zone thickness is $0.02$ in the first step. In step 75, at $X_1=0$ and $X_2=0.5$, the damage zone thickness has increased to $0.2$, which is 10 times greater than the initial thickness. This substantial increase poses a major challenge in GED modeling, hindering us from achieving accurate simulations.

\begin{figure}[h!]
    \centering
    \includegraphics[width=0.6\linewidth]{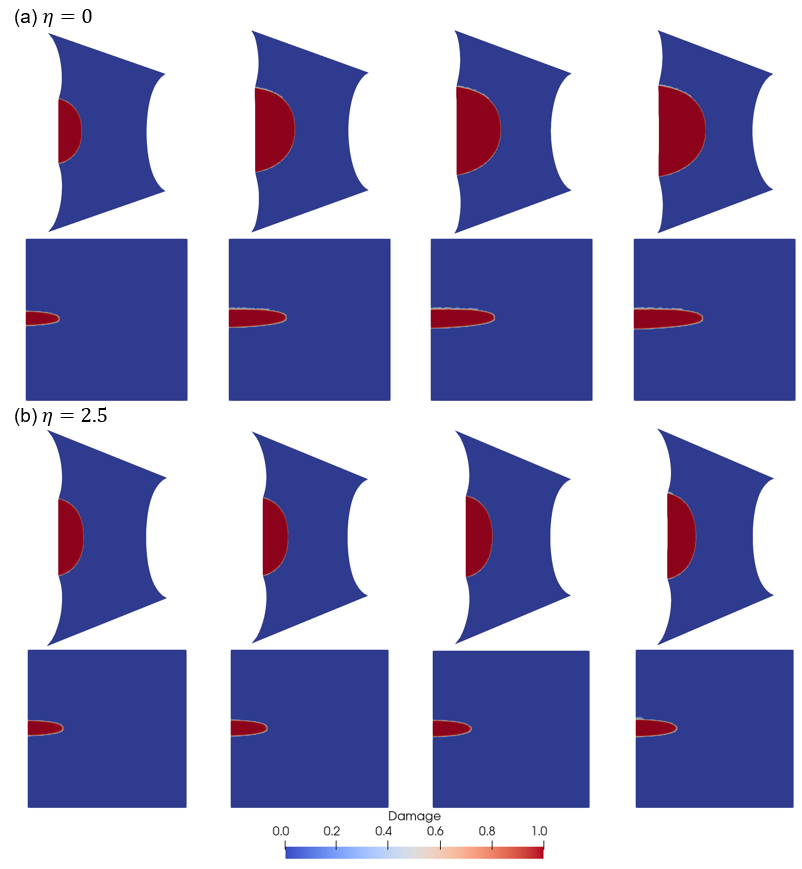}
    \caption{The effect of artificial viscosity on the crack propagation in GED modeling for four consecutive time steps after the onset of crack propagation for an edge-crack sample under triangular-type loading. Note that at the initial state, the crack is represented diffusely.}
    \label{fig:example6-eta-contour}
\end{figure}

To gauge the impact of artificial viscosity on GED versus phase-field simulations, we varied artificial viscosity as follows: $\eta = \left\lbrace 0, 1, 2.5, 5\right\rbrace$. Fig. \ref{fig:example6-eta-contour} depicts the contours for $\eta=0$ and $\eta=2.5$, showcasing a noticeable slowdown in crack propagation with the addition of artificial viscosity. This slowdown in crack propagation is possibly more pronounced than that observed in the phase-field simulations.

\begin{figure}[hbt!]
    \centering
    \includegraphics[width=0.87\linewidth]{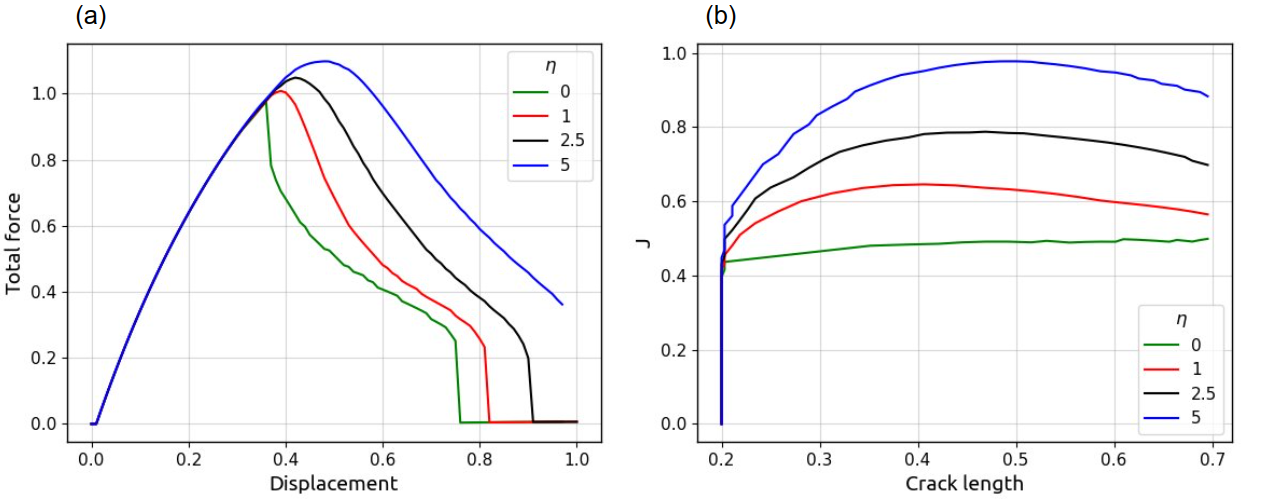}
    \caption{The effect of artificial viscosity parameter on the energy release rate for an edge-crack sample under triangular-type loading utilizing the GED framework.}
    \label{fig:example6-graph}
\end{figure}

In this regard, Fig. \ref{fig:example6-graph} provides valuable insights. Firstly, in Fig. \ref{fig:example6-graph}(b) for the $\eta=0$ case, we observe that J-integral approaches $0.5$ at crack initiation, but monotonically increases during propagation. This mild increase of the predicted energy release rate is a downstream effect of the damage zone broadening phenomenon. Additionally, increasing the artificial viscosity parameter from $0$ to $5$ raises the peak of J-integral from $0.5$ to $0.98$, representing a substantial $96\%$ increase. In contrast, as shown in Fig. \ref{fig:example3-graph}, a similar artificial viscosity parameter change in phase-field simulations led to a significantly less pronounced $15\%$ increase. 
The underlying reason is that in the phase-field method, artificial viscosity is related to changes in the damage field near the crack tip, while in the GED formulation, artificial viscosity pertains to variations in the stretch field across the entire domain, resulting in a more significant enhancement to the energy release rate.

\section{Conclusion}\label{Section:conclusion}

In this study, we delved into the verification of crack propagation modeling approaches as they pertain to elastomers. The foundation of the models presented was a near-incompressible hyperelastic material Neo-Hookean strain energy. Our approach began with utilizing the phase-field method to explore solution convergence under an unstable loading scenario. To mitigate instability, we introduced artificial viscosity and analyzed its impact on the energy release rate, a task that had not been previously tested for finite deformation scenarios and at the limit of near-incompressibility. We examined predictions for the energy release rate during propagation and adapted a domain J-integral approach for this task. Subsequently, we introduced a novel stretch-based gradient-enhanced damage (GED) model to simulate crack evolution, providing a comparative analysis with the phase-field method. Key findings from our study include:

\begin{itemize}
  \item For phase-field and GED simulations for fracture of elastomeric materials, artificial viscosity not only delays crack initiation but also significantly alters the fracture energy prediction. Notably, its impact is most pronounced during the initial stages of crack propagation.
  \item Analyzing both diffuse crack and discrete pre-crack representations demonstrated significant differences in system response and fracture energy predictions near crack initiation.
  \item The GED method was successful at concentrating "crack-like" features at finite strains and at the limit of near incompressibility.
  \item The GED method provided reasonable fracture energy predictions. However, it has been demonstrated that similar to the strain-based GED model for linear problems, the stretch-based model also suffers from damage-zone broadening (a problem not present in the phase-field method).
  \item The investigation revealed that the impact of artificial viscosity on GED is substantially more pronounced compared to the phase-field method.
  
\end{itemize}

In future work, a compelling avenue for exploration could involve addressing the broadening issue inherent in the GED model for strain- and stretch-based formulations, including a connection to statistical mechanics-motivated constitutive models for elastomers. This aspect presents an intriguing challenge and offers opportunities for refining the model's accuracy and effectiveness in simulating brittle crack evolution in elastomeric materials. In this way, physically diffuse information such as diffuse chain scission could possibly be captured by such a refined model.

\section{Acknowledgments}

The authors are grateful to Dr. Brandon Talamini for useful discussions. SMM and NB acknowledge the support of the National Science Foundation under Grant No. CMMI-2038057. JPM gratefully acknowledges the support of the National Science Foundation Graduate Research Fellowship Program under Grant No. DGE-1650441. Any opinions, findings, conclusions, or recommendations expressed in this material are those of the author(s) and do not necessarily reflect the views of the National Science Foundation. JPM also gratefully acknowledges the support of UES, Inc. (a BlueHalo Company), the Air Force Research Laboratory Materials and Manufacturing Directorate, and the National Research Council (NRC) Research Associateship Program (administered by the National Academies of Sciences, Engineering, and Medicine).

\bibliographystyle{elsarticle-num}

\end{document}